\documentclass[aps,prl,twocolumn,groupedaddress,showpacs,onlinecite]{revtex4-1}

\usepackage{color}
\usepackage{graphicx}
\usepackage{dcolumn}
\usepackage{bm}
\usepackage{float}
\usepackage[mathlines]{lineno}
\usepackage[colorlinks,linkcolor=blue,anchorcolor=blue,citecolor=blue,urlcolor=blue,hyperindex,CJKbookmarks,dvipdfm]{hyperref}

\usepackage{amsmath}
\usepackage{txfonts}

\begin{document}

\title{Electron-phonon coupling in a honeycomb borophene grown on Al(111) surface}

\author{Miao Gao$^{1}$}\email{gaomiao@nbu.edu.cn}
\author{Xun-Wang Yan$^{2}$}
\author{Jun Wang$^{1}$}
\author{Zhong-Yi Lu$^{3}$}
\author{Tao Xiang$^{4,5}$}

\date{\today}

\affiliation{$^{1}$Department of Microelectronics Science and Engineering, School of Physical Science and Technology, Ningbo University, Zhejiang 315211, China}

\affiliation{$^{2}$College of Physics and Engineering, Qufu Normal University, Shandong 273165, China}

\affiliation{$^{3}$Department of Physics, Renmin University of China, Beijing 100872, China}

\affiliation{$^{4}$Institute of Physics, Chinese Academy of Sciences, Beijing 100190, China }

\affiliation{$^{5}$School of Physics, University of Chinese Academy of Sciences}

\begin{abstract}
  Recently, a honeycomb borophene was reported to grow successfully on Al(111) surface.
  Since the metallic $\sigma$-bonding bands of honeycomb boron sheet play
  a crucial role in the 39 K superconductivity of MgB$_2$, it is physically interesting
  to examine whether similar property exists in this material.
  We have calculated the electronic structures and the electron-phonon coupling
  for honeycomb borophene by explicitly considering the substrate effect
  using first-principles density functional theory in conjunction with
  the Wannier interpolation technique.
  We find that the $sp^2$-hybridized $\sigma$-bonding bands of honeycomb
  borophene are metallized due to moderate charge transfer from the Al substrate,
  similar as in MgB$_2$.
  However, the electron-phonon coupling in honeycomb borophene is much weaker than in MgB$_2$ due to the hardening
  of the bond-stretching boron phonon modes and the reduction of phonon density of states.
  Nevertheless, the interlayer coupling between Al-associated phonons and electrons in
  borophene is strong.
  Based on this observation, we predict that a 6.5 K superconducting transition
  can be observed in a free-standing borophene decorated by a single Al layer, namely monolayer AlB$_2$.
  Accordingly, similar superconducting transition temperature could be expected in honeycomb borophene on Al(111).
\end{abstract}

\maketitle

\section{I.  Introduction}

Boron possesses rich chemistry like carbon.
Since the discovery of graphene, great effort has been made to synthesize two-dimensional (2D) boron sheet, namely borophene.
In sharp contrast to graphene, it was suggested from theoretical calculations that a borophene favors a triangular lattice structure with a special vacancy order, exemplified by the so-called $\alpha$- or $\beta$-sheet structured borophene \cite{Yang-PRB77,Penev-NanoLett12,Yu-JPCC116,Tang-PRB82,Wu-ACSNano6}.
Recently, three different structured borophene were successfully grown on Ag(111) surface by direct evaporation of boron atoms \cite{Mannix-Science,Feng-arXiv1}.
One has a buckled structure ($b$-B$_2$) without vacancies, while the other two have vacancy orders, named as $\beta_{12}$-B$_5$ and $\chi_3$-B$_4$, respectively.

Searching for superconductivity in 2D compounds is of great interest, because it has potential application in
constructing nanoscale superconducting devices with single-electron
sensitivity \cite{Franceschi-Nature5,Huefner-PRB79}.
In addition to graphene and borophene, phosphorene \cite{Liu-Nano8,Li-Nature9}, silicene \cite{Lalmi-APL97,Chen-PRL109}, germanene \cite{Davila-NJP16,Derivaz-NanoLett15}, and stanene \cite{Zhu-NatMat14,Zang-arXiv}
have all been synthesized in laboratory.
Phosphorene is a semiconductor with a direct bandgap of 1.51 eV \cite{Qiao-NatCommun5}.
Silicene, germanene, and stanene are zero-gap semimetals with vanished density of states at the Fermi energy \cite{Cahangirov-PRL102,Xu-PRL111},
similar to graphene.
As a result, additional charge doping must be introduced to induce superconductivity.
In recent years, a number of theoretical calculations have been done to explore the possibility of superconductivity with high transition temperature in these doped 2D materials \cite{Profeta-Nature8,Wan-EPL104,Shao-EPL108,Ge-NJP17,Si-PRL111}.
More specifically, a superconducting transition was predicted to occur at a temperature as high as 31.6 K in heavily electron-doped graphene \cite{Si-PRL111}.
Similar prediction was made for arsenene \cite{Kong-CPB27}.
However, superconductivity was only observed in Li-intercalated
few-layer graphene at 7.4 K \cite{Tiwari-arXiv}, in magic-angle graphene superlattice at 1.7 K \cite{Cao-Nature556}, and in bilayer stanene below 1 K \cite{Liao-arXiv}.
This suggests that it is difficult to dope charge into these systems up to the level assumed in theoretical calculations.

Different from these zero-gap or gapped 2D compounds,
borophenes, including $b$-B$_2$, $\beta_{12}$-B$_5$, $\chi_3$-B$_4$, are all intrinsic metals \cite{Mannix-Science,Feng-arXiv1}.
They are good candidates of intrinsic 2D superconductors.
From first-principles electronic structure calculations for the electron-phonon coupling (EPC), it was predicted that these three free-standing borophenes could become superconducting around 20 K even without doping \cite{Penev-Nano16,Zhao-APL108,Gao-PRB95,Xiao-APL109}.
However, the tensile strain with the electron transfer imposed by the Ag substrate can significantly suppress the superconducting transition temperature $T_c$.
For example, the $T_c$ of $\beta_{12}$-B$_5$ could be dramatically reduced to 0.09 K by a biaxial tensile strain of 2\% with an electron transfer of 0.1 $e$/boron \cite{Cheng-2D4}.
Thus, it is indispensable to include the substrate effect in the calculation of EPC in borophenes and other 2D materials.

Layered transition-metal dichalcogenide (TMD) is another family, whereas superconductivity has been extensively studied.
For semiconducting TMDs ($2H$-MoS$_2$, $2H$-MoSe$_2$, $2H$-MoTe$_2$, $2H$-WS$_2$, $1T$-TiSe$_2$, and $1T$-TaS$_2$), superconductivity can be induced by applying chemical intercalation \cite{Somoano-PRL27,Morosan-NP2}, electrostatic gating \cite{Li-Nature529,Yu-NN10,Ye-Science338,Costanzo-NN11,Shi-SR5}, or pressure \cite{Chi-PRL120}.
Metallic TMDs, including $2H$-type NbSe$_2$, TaS$_2$, and TaSe$_2$, show intrinsic superconductivity, coexisting with a charge-density wave (CDW) order at low temperatures \cite{Revolinsky-JPCS26,Valla-PRL85,Nagata-JPCS53,Freitas-PRB93}.
Recently, lots of attentions have been paid to monolayer TMDs to investigate the effect of
dimensionality on both superconductivity and CDW.
For example, bulk $2H$-TaS$_2$ enters the CDW state at 70 K, and becomes a superconductor at 0.8 K \cite{Nagata-JPCS53,Castro-PRL86,Guillamon-NJP13,Navarro-NC7}.
The $T_c$ of $2H$-TaS$_2$ is firstly improved to 2.2 K with the thickness down to five covalent planes \cite{Navarro-NC7},
and then to 3.4 K in the monolayer limit \cite{Yang-PRB98}. In sharp contrast to the boosted $T_c$, the CDW state vanishes in monolayer $2H$-TaS$_2$ \cite{Yang-PRB98}.
But these trends of superconductivity and CDW in approaching the 2D limit are not universal in TMDs. Specifically, the $T_c$ of $2H$-NbSe$_2$ decreases from 7.2 K in bulk crystal to 3 K in monolayer \cite{Xi-NP12},
where the CDW order is strongly enhanced \cite{Xi-NN10}.
For 2D intrinsic superconductors, monolayer TMD and borophene are somehow complementary.
On one hand, although stable monolayer TMD can be easily exfoliated from the bulk phase, the $T_c$ is relatively lower.
On the other hand, free-standing borophene can have a high $T_c$, but current structures of borophene can only exist on special substrate, e.g. Ag.
Moreover, the interplay between borophene and Ag substrate has an inhibited effect on superconductivity.

Recently, a honeycomb borophene ($h$-B$_2$) was successfully grown on the Al(111) substrate  (abbreviated as $h$-B$_2$/Al(111) hereafter) \cite{Li-arXiv1712}.
From first-principles electronic structure calculations, it was found that the honeycomb structure is stabilized by a significant charge transfer from Al(111) to $h$-B$_2$ \cite{Li-arXiv1712}.
Similar as in MgB$_2$ whose honeycomb boron sheet plays a central role in pairing electrons in a relatively high $T_c$ \cite{MgB2},
a free-standing $h$-B$_2$ was also predicted to be a 30 K superconductor by ignoring the imaginary phonon modes \cite{Zhao-APL108}.
However, it is not clear whether such high-$T_c$ superconductivity can survive in $h$-B$_2$ grown on an Al(111) substrate.

In this paper,
we explore the possible superconductivity in $h$-B$_2$/Al(111) by calculating its electronic structure, lattice dynamics, and EPC
using the first-principles density functional theory and the Wannier interpolation technique.
The influence of the Al substrate is fully considered in our calculations,
which allows both the charge transfer and strain effects to be more preciously determined.
After full optimization, we find that $h$-B$_2$/Al(111) has a buckling lattice structure
with a tiny buckling height of 0.035 {\AA}.
The $sp^2$-hybridized $\sigma$-bonding bands of $h$-B$_2$/Al(111) are partially filled,
but the intra-boron-layer EPC in this material is weak due to the hardening of phonons
and the reduced phonon DOS.
However, there is a strong coupling between Al phonons and electrons
in $h$-B$_2$.
Based on this property, we propose a monolayer AlB$_2$ material with an EPC constant $\lambda$
about 35\% larger than in MgB$_2$ \cite{Margine-PRB87}.
Using the McMillian-Allen-Dynes formula, we predict that the superconducting $T_c$ of this
monolayer AlB$_2$ is about 6.5 K.

\section{II.  Computational method}

In the calculations, the first-principles package, Quantum-ESPRESSO, was adopted \cite{pwscf}. We
calculated the Bloch states and the phonon perturbation potentials \cite%
{Giustino-PRB76} using the local density approximation and the
norm-conserving pseudopotentials.
The kinetic energy
cut-off and the charge density cut-off were taken to be 80 Ry and 320 Ry,
respectively.
The charge densities were calculated on an unshifted mesh of 40$\times $40$\times $1 points in combination with
a Methfessel-Paxton smearing \cite{Methfessel-PRB40} of 0.02 Ry. The dynamical matrices and the perturbation potentials
were calculated on a $\Gamma $-centered 10$\times $10$\times $1 mesh, within
the framework of density-functional perturbation theory \cite{Baroni-RMP73_515}.

\begin{figure}[tbh]
\begin{center}
\includegraphics[width=8.6cm]{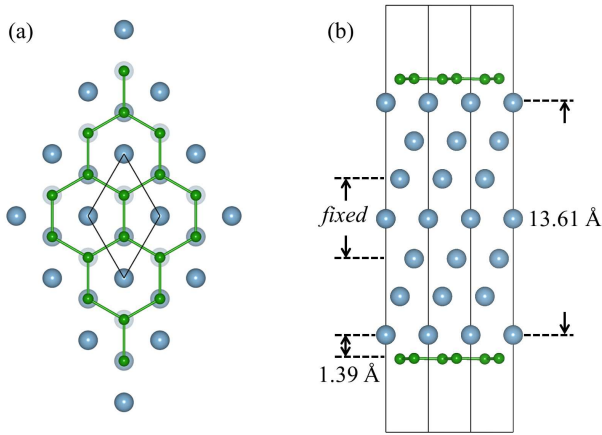}
\caption{Slab model of $h$-B$_2$/Al(111). (a) Top view plotted with the depth-cueing technique. (b) Side view. The green and slate balls represent boron and Al atoms, respectively. The thick black line denotes the unit cell. }
\label{fig:Structure}
\end{center}
\end{figure}

In the construction of slab model, if only one borophene layer is deposited on Al surface, another surface of Al, namely fracture surface, will be exposed to the vacuum (see Appendix A for the results calculated with semi-infinite slab).
To avoid the influence of Al surface states on the fracture surface,
we employed a slab model with inversion symmetry to simulate $h$-B$_2$/Al(111)
[Fig.~\ref{fig:Structure}].
There are seven Al layers with one $h$-B$_2$ on each surface.
The thickness of vacuum layer in the slab was set to 8.58 {\AA}.
The experimental value of the in-plane lattice constant of the Al(111) surface, 2.8635 {\AA}
\cite{Wyckoff}, was adopted.
Al atoms in the middle three layers were fixed to mimic the bulk,
the other Al layers and two $h$-B$_2$ sheets were relaxed to minimize the total energy.

The maximally localized Wannier functions (MLWFs) \cite%
{Marzari-PRB56,Souza-PRB65} were constructed on a 10$\times $10$\times $1
grid of the Brillouin zone.
We used 38 Wannier functions to describe the band structure of $h$-B$_2$/Al(111) around the Fermi level.
Among them, four are $p_z$-like states associated with the boron atoms, six are $\sigma$-like
states localized in the middle of boron-boron bonds, the other 28 functions correspond to the $s$ and $p$ orbitals of seven Al atoms.
These MLWFs are well localized in space. For instance, the average spatial spread of the six $\sigma$-like states is just about 1.01 {\AA}$^2$.
Fine electron (360$\times $360$\times $1) and phonon (120$%
\times $120$\times $1) grids were used to interpolate the EPC constant through
the Wannier90 and EPW codes \cite{Mostofi-CPC178,Noffsinger-CPC181}. The Dirac $%
\delta $-functions for electrons and phonons were smeared out by a Gaussian function with the
widths of 50 meV and 0.2 meV, respectively.

The EPC constant $\lambda $ was determined by the
summation of
the momentum-dependent coupling constant $\lambda _{\mathbf{q}\nu }$
over the first Brillouin zone, or the integration of the Eliashberg
spectral function $\alpha ^{2}F(\omega )$ \cite{Allen-PRB6_2577,Allen-RPB12_905},
\begin{equation}
\label{eq:lambda}
\lambda=\frac{1}{N_{\bf q}}\sum_{{\bf q}\nu}\lambda_{{\bf q}\nu}=2\int\frac{\alpha^2F(\omega)}{\omega}d\omega.
\end{equation}
The coupling constant $\lambda_{\mathbf{q}\nu }$ reads
\begin{equation}
\lambda _{\mathbf{q}\nu }=\frac{2}{\hbar N(0)N_{\bf k}}\sum_{nm\mathbf{k}}\frac{1%
}{\omega _{\mathbf{q}\nu }}|g_{\mathbf{k},\mathbf{q}\nu }^{nm}|^{2}\delta
(\epsilon _{\mathbf{k}}^{n})\delta (\epsilon _{\mathbf{k+q}}^{m}).
\label{eq:lambda_qv}
\end{equation}%
$N_{\bf q}$ and $N_{\bf k}$ represent the total numbers of \textbf{q} and \textbf{k} points in the fine
\textbf{q}-mesh and \textbf{k}-mesh, respectively.
Similarly, we can also define a {\bf k}- and band-resolved coupling constant $\lambda_{{\bf k}n}$ \cite{Margine-PRB87,Margine-PRB90,Penev-Nano16}.
\begin{equation}
\lambda_{{\bf k}n}=\frac{2}{\hbar N(0)N_{\bf q}}\sum_{\nu m\mathbf{q}}\frac{1}{\omega_{{\bf q}\nu}}|g_{{\bf k},{\bf q}\nu}^{nm}|^2\delta(\epsilon_{{\bf k}}^{n})\delta(\epsilon_{{\bf k+q}}^{m}).
\end{equation}
From $\lambda_{\mathbf{q}\nu }$ or $\lambda_{{\bf k}n}$, we will identify
which phonon mode or electronic state has larger contribution to the EPC.
The Eliashberg spectral function $\alpha^2F(\omega)$ was calculated with
\begin{equation}
\alpha^2F(\omega)=\frac{1}{\hbar N(0)N_{\bf k}N_{\bf q}}\sum_{\nu mn\mathbf{kq}}|g_{{\bf k},{\bf q}\nu}^{nm}|^2\delta(\epsilon_{{\bf k}}^{n})\delta(\epsilon_{{\bf k+q}}^{m})\delta(\omega-\omega_{{\bf q}\nu}).
\end{equation}
Here $\omega _{\mathbf{q}\nu }$ is the phonon frequency and $g_{\mathbf{k},%
\mathbf{q}\nu }^{nm}$ is the probability amplitude for scattering an
electron with a transfer of crystal momentum $\mathbf{q}$. $\left(
n, m\right) $ and $\nu $ denote the indices of energy bands and phonon mode,
respectively. $\epsilon _{\mathbf{k}}^{n}$ and $\epsilon _{\mathbf{k+q}}^{m}$
are the eigenvalues of the Kohn-Sham orbitals with respect to the Fermi
level. $N(0)$ is the density of states (DOS) of electrons at the Fermi
level.
It is noted that above formulas can not be directly used to evaluate the intrinsic EPC of $h$-B$_2$,
since there is significant $N(0)$ from the Al substrate. We will present the strategy about how to extract the EPC of $h$-B$_2$ from slab calculation in section IV.
This extraction is based on a fact that the boron and Al phonons are separated in frequency.

The superconducting transition temperature is determined by the McMillian-Allen-Dynes formula \cite{Allen-RPB12_905},
\begin{equation}
T_{c}=\frac{\omega _{\text{log}}}{1.2}\exp \left[ \frac{%
-1.04(1+\lambda )}{\lambda (1-0.62\mu ^{\ast })-\mu ^{\ast }}\right],
\label{eq:Tc}
\end{equation}%
in which $\mu ^{\ast }$ is the Coulomb pseudopotential, $\omega _{\text{log}}$ is the logarithmic average frequency,
\begin{equation}
\omega _{\text{log}}=\exp \left[ \frac{2}{\lambda }\int \frac{d\omega }{%
\omega }\alpha ^{2}F(\omega )\ln \omega \right].
\end{equation}

\section{III.  Results and analysis}

From the calculation, we find that the crystal structure of $h$-B$_2$/Al(111), with boron atoms occupying the hollow sites of the triangular lattice of outmost Al layer [Fig.~\ref{fig:Structure}(a)], is energetically favored.
After optimization,
the total thickness of Al substrate and the average distance of boron atoms from the outmost Al layer are 13.61 {\AA} and 1.39 {\AA}, respectively.
The $h$-B$_2$ layer is nearly flat with a tiny buckling of 0.035 {\AA}, not reported in the previous calculation \cite{Li-arXiv1712}.
This buckling structure is robust.
It exists even when the vacuum layer is increased to 23.58 {\AA}.

\begin{figure}[tbh]
\begin{center}
\includegraphics[width=8.6cm]{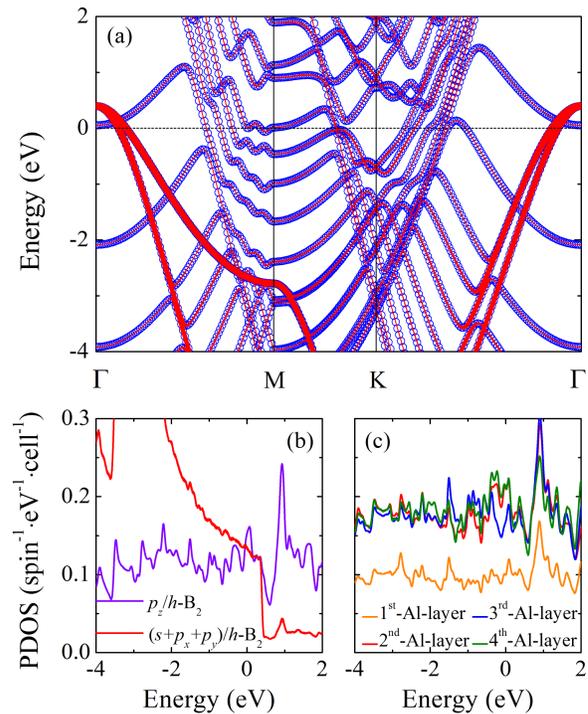}
\caption{(a) Band structures of $h$-B$_2$/Al(111).
The width of red lines is proportional to the contribution of the $sp^2$-hybridized $\sigma$-bonding orbitals to the Kohn-Sham states.
The blue circles are obtained by interpolation of MLWFs. (b) Projected density of states (PDOS) of boron orbitals.
(c) PDOS of Al layers. The Al layers are numbered according to their distance from borophene. For instance, the 1$^{st}$-Al-layer denotes the closest Al layer to borophene, i.e., the outmost Al layer. The Fermi level is set to zero. }
\label{fig:Band}
\end{center}
\end{figure}

Figure \ref{fig:Band} shows the band structures and projected density of states (PDOS) of $h$-B$_2$/Al(111).
There are nine bands across the Fermi level [Fig.~\ref{fig:Band}(a)]. The MLWFs interpolated band structure agrees excellently with the first-principles calculation.
This, together with the high locality of MLWFs, sets a solid foundation for accurately computing the EPC using the Wannier interpolation technique.
The band structure suggests that the B-B $sp^2$-hybridized $\sigma$-bonding bands are partially occupied, similar as in MgB$_2$, but the
$\sigma$-bonding bands in the bulk AlB$_2$ are completely filled \cite{Medvedeva-PRB64}.
By calculating the Bader charge decomposition \cite{Henkelman-CMS36,Tang-JPCM21}, we find that in $h$-B$_2$/Al(111), the charge transfer to the inward (outward) boron atom is 1.15 (0.87) $e$/boron, contributed mainly by the outmost Al atom.
While in the bulk AlB$_2$, each boron atom acquires about 1.46 electrons.
At the Fermi level, the $\sigma$-like and the $p_z$ orbitals of boron have almost
equal proportion in PDOS [Fig.~\ref{fig:Band}(b)]. More importantly, Al layers dominate the contribution to $N(0)$ [Fig.~\ref{fig:Band}(c)].
The relatively smaller occupation number in electronic states associated with the outmost Al layer is consistent with the charge transfer picture.
For the inner Al layers, the PDOS curves merge together [Fig.~\ref{fig:Band}(c)], suggesting that the interaction between borophene and inner Al layers is rather weak.

\begin{figure}[tbh]
\begin{center}
\includegraphics[width=8.6cm]{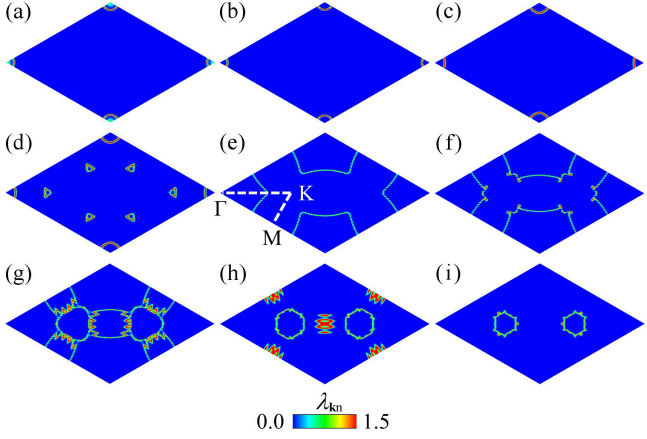}
\caption{Fermi surfaces formed by the nine bands across the Fermi level in $h$-B$_2$/Al(111) in the
reciprocal unit cell. Different colors represent the strength of $\lambda_{{\bf k}n}$.
We label these nine Fermi surfaces as FS-$\alpha$, with
$\alpha$ running from $a$ to $i$.}
\label{fig:FS}
\end{center}
\end{figure}

Figure \ref{fig:FS} shows the Fermi-surface contours for the nine bands crossing the Fermi level.
By projecting $\lambda_{{\bf k}n}$ onto these Fermi surfaces, we find that
electrons on the Fermi surfaces around the $\Gamma$  [Fig.~\ref{fig:FS}(a)-Fig.~\ref{fig:FS}(d)]
and $M$ [Fig.~\ref{fig:FS}(h)] points couple strongly with phonons.
From Fig.~\ref{fig:Band}, we know that the $\Gamma$-centered Fermi sheets stem from the $sp^2$-hybridized B-B $\sigma$-bonding bands.
The orbitals of B-$p_z$ and Al comprise the electronic states around the $M$ point on FS-$h$.
Compared with the outward boron atoms, there are more electrons transferred to the inward ones, causing the $p_z$ orbitals more insulating, especially around the $M$ point.

\begin{figure}[tbh]
\begin{center}
\includegraphics[width=8.6cm]{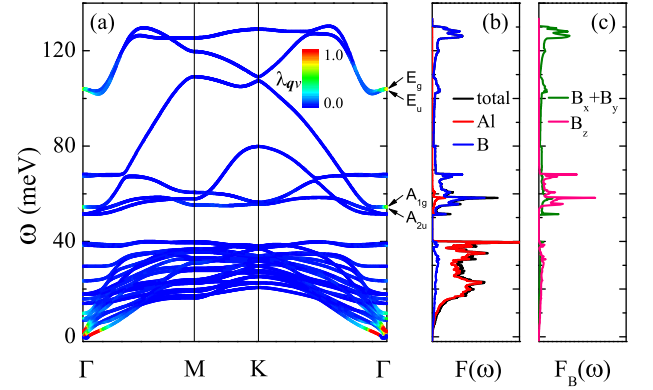}
\caption{Lattice dynamics of $h$-B$_2$/Al(111).
(a) Phonon spectrum with a color representation of $\lambda_{{\mathbf q}\nu}$ at a given wave vector and mode. The mode symmetries of strongly coupled phonon modes at $\Gamma$ are labelled on the right side. (b) Projected phonon density of states. (c) Decomposed boron-related phonon density of states along three directions.}
\label{fig:phonon}
\end{center}
\end{figure}

Figure \ref{fig:phonon} shows the lattice dynamics of $h$-B$_2$/Al(111).
A gap of about 10 meV is found in the phonon spectrum [Fig.~\ref{fig:phonon}(a)].
Due to the large difference between the masses of boron and Al atoms, it is straightforward to assign the modes above and below the gap mainly to boron and Al atoms, respectively.
This assumption is confirmed by the result of projected phonon DOS calculated under the quasi-harmonic approximation [Fig.~\ref{fig:phonon}(b)].
The phonons at the $\Gamma$ point have the largest coupling with electrons.
The strongly coupled modes of $h$-B$_2$ at the $\Gamma$ point
are identified to be $A_{2u}$, $A_{1g}$, $E_u$, and $E_g$ [Fig.~\ref{fig:phonon}(a)].
The phonon DOS attributed to boron atoms can be classified into two regions [Fig.~\ref{fig:phonon}(c)].
From 50 meV to 70 meV, the phonons originate from the out-of-plane movements of boron atoms.
The in-plane displacements of boron atoms mainly participate in the phonons above 100 meV.
This indicates that the strongly coupled $E_u$ and $E_g$ modes at about 104 meV are mainly the contribution of the in-plane bond-stretching phonon modes.

\begin{figure}[tbh]
\begin{center}
\includegraphics[width=8.6cm]{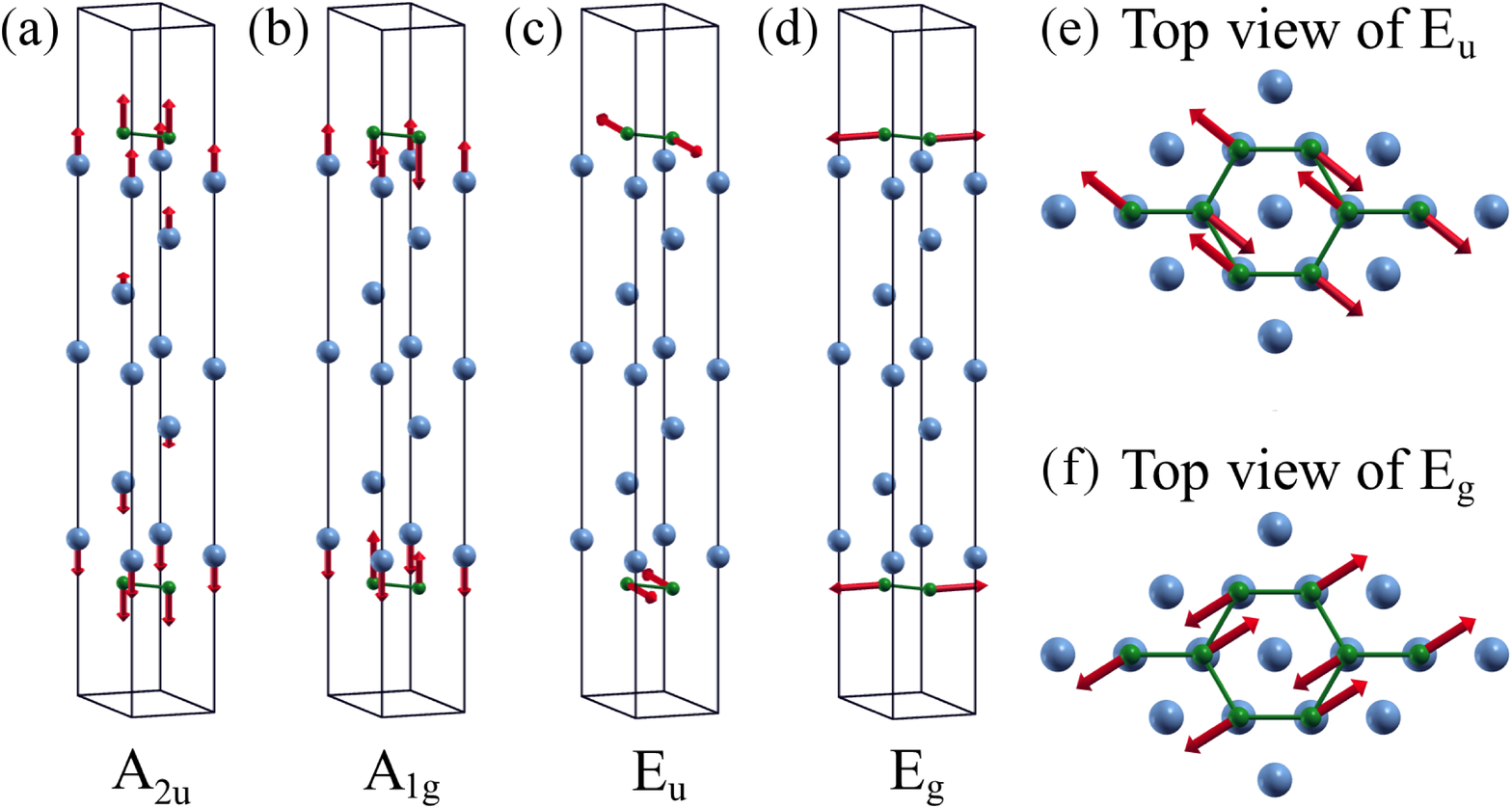}
\caption{Vibrational patterns for strongly coupled phonon modes of $h$-B$_2$ at the $\Gamma$ point.
The red arrows and their lengths denote the directions and relative amplitudes of atomic movements, respectively.}
\label{fig:mode}
\end{center}
\end{figure}

The vibrational patterns of strongly coupled phonons at the $\Gamma$ point above the frequency gap are schematically shown in Fig.~\ref{fig:mode}.
The amplitudes of Al displacements in the strongly coupled phonon modes gradually decrease with the increase of vibrational frequency. This agrees with the fact that the frequencies of the Al phonon modes lie mainly below 40 meV.
Particularly, the high-frequency $E_u$ and $E_g$ phonons only involve
the 2D bond-stretching motions of boron atoms [Fig.~\ref{fig:mode}(c)-Fig.~\ref{fig:mode}(f)], consistent with the phonon DOS of boron atoms shown in Fig.~\ref{fig:phonon}(c).

The Eliashberg spectral function $\alpha^2F(\omega)$ is divided into three regions  separated by two gaps [Fig.~\ref{fig:a2F}(a)].
A large proportion of $\alpha^2F(\omega)$ lies below 40 meV. This indicates that the Al-associated phonons play a vital role in the EPC of $h$-B$_2$/Al(111).
With respect to $F(\omega)$ [Fig.~\ref{fig:phonon}(b)], the sharp peaks of $\alpha^2F(\omega)$ around 55 meV and 104 meV show that there exist strongly coupled
B-associated phonons, such as the $A_{2u}$, $A_{1g}$, $E_u$, and $E_g$ modes [Fig.~\ref{fig:mode}].
The EPC constant $\lambda$ and $\omega_\text{log}$ are found to be 0.64 and 22.15 meV.
The Al-associated and B-associated phonons contribute 76.6\% and 23.4\% to the total $\lambda$, respectively.
We also find that there is a strong interfacial coupling between Al phonons and electrons in $h$-B$_2$. It occupies about 30.7\% of the total $\lambda$.

\section{IV.  Discussion}

The Eliashberg formula was established for a superconductor with translational symmetry.
Attention, however, should be paid in the calculation of EPC in 2D compounds using this formula when a metallic substrate is included.
In particular, Al is a good metal with a large bandwidth. More Al-associated electronic bands would appear around the Fermi level if the thickness of Al substrate gradually increases in the slab model that we used to simulate $h$-B$_2$/Al(111).
The contribution of $h$-B$_2$ to the EPC will be submerged by the significant reduction of boron atoms to the DOS around the Fermi level.
As a result, the EPC of $h$-B$_2$/Al(111) supercell will approach to the EPC of bulk face-centered-cubic (fcc) Al \cite{Savrasov-PRB54},
when the thickness of substrate becomes infinite in the slab model.
Thus the intrinsic EPC of $h$-B$_2$ need to
be extracted reasonablely from current data. Moreover, the previously obtained $\lambda$ (0.64) and $\omega_{\text{log}}$ (22.15 meV) can not be
substituted directly into the McMillian-Allen-Dynes formula to determine the $T_c$ of $h$-B$_2$/Al(111).

\begin{figure}[tbh]
\begin{center}
\includegraphics[width=8.6cm]{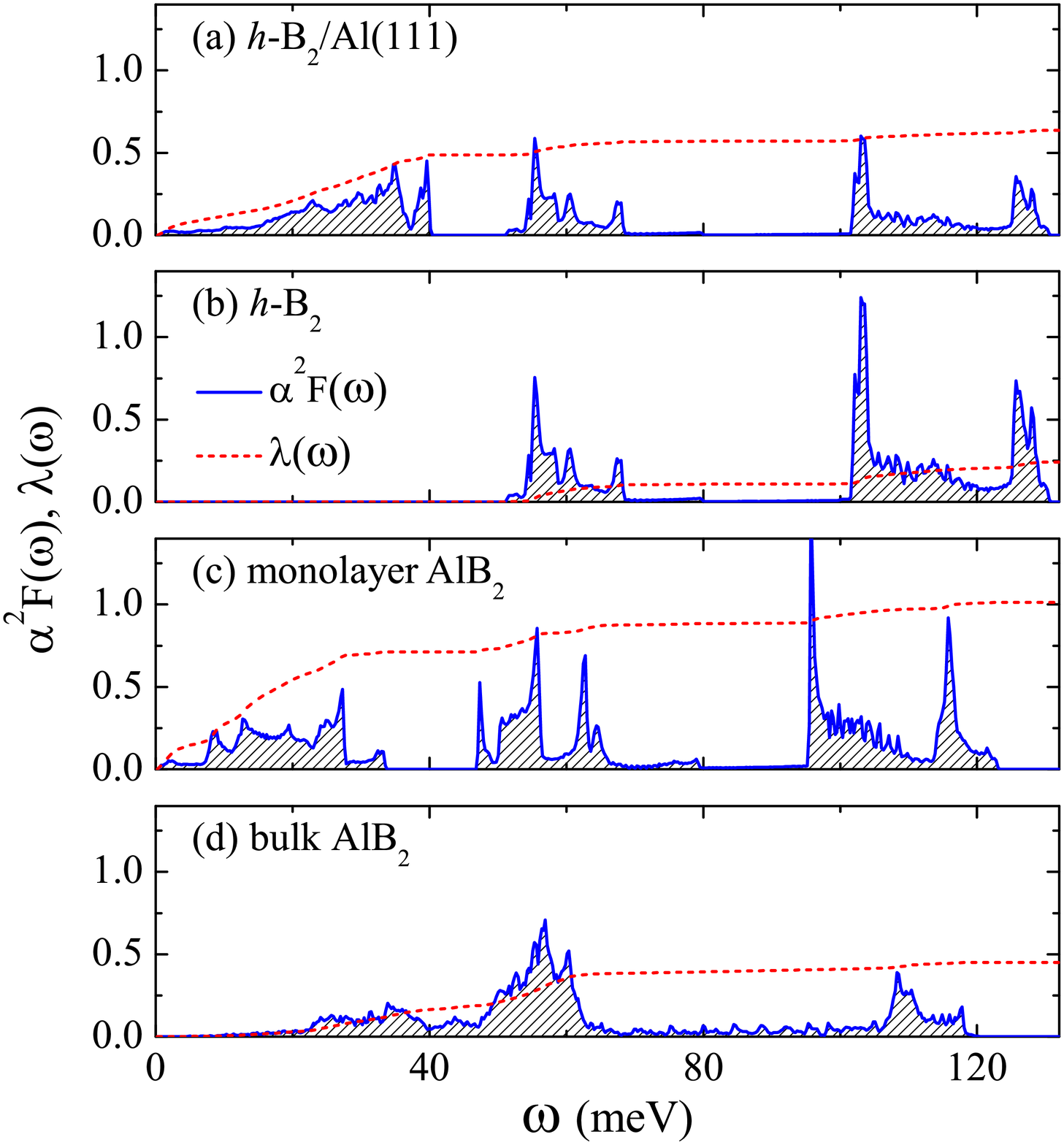}
\caption{Eliashberg spectral function $\alpha^2F(\omega)$ for (a) $h$-B$_2$/Al(111), (b) $h$-B$_2$, (c) monolayer AlB$_2$, and (d) bulk AlB$_2$.
The accumulated $\lambda(\omega)$ is computed using the formula $2\int_0^\omega \frac{1}{\omega'}\alpha^2F(\omega')d\omega'$.}
\label{fig:a2F}
\end{center}
\end{figure}

Since the boron and Al phonon vibrations are well separated in frequency, we can regard the EPC generated by boron phonons as the EPC of $h$-B$_2$ if the interfacial coupling between the boron phonons and the electronic states from the substrate can be properly deducted.
We calculated $\lambda_{{\bf k}n}$ in the frequency region from 50 to 140 meV.
It is found that the $\sigma$-bonding electrons of $h$-B$_2$ contribute 39.1\% (62.6\%) to the EPC caused by the out-of-plane (in-plane) boron phonons.
So the $\alpha^2F(\omega)$ in the intervals [50, 100] meV and [100, 140] meV should be reduced by 60.9\% and 37.4\%, respectively, if we assume $\alpha^2F(\omega)$ depends weakly on $\omega$.
Furthermore, the EPC constant can not be normalized by the whole DOS of the $h$-B$_2$/Al(111) supercell at the Fermi level [see Eq.~\eqref{eq:lambda_qv}].
Instead, it should be normalized only by the electron DOS of $h$-B$_2$.
At the Fermi level, the total DOS and the DOS contributed by the $h$-B$_2$ electrons are 1.91 and 0.58 states/spin/eV/cell, respectively.
From the refined $\alpha^2F(\omega)$ of $h$-B$_2$, which is shown in Fig.~\ref{fig:a2F}(b), we find the value of $\lambda$ for $h$-B$_2$ is only 0.24.
Additionally, we can also roughly estimate the EPC for Al substrate following above procedure. After deducting the electronic states of boron from Fig.~\ref{fig:FS}, we find that the remainder EPC strength below the frequency gap is about 0.33.
Further scaling with the Al-associated $N(0)$, the EPC constant for Al substrate is determined to be 0.54, close to 0.44 calculated previously \cite{Savrasov-PRB54}.
This is probably because the inner five Al layers still resemble the bulk Al [see Fig.~\ref{fig:Structure}(b) and Fig.~\ref{fig:Band}(c)].

The usage of inversion slab with two $h$-B$_2$ sheets may raise the question about the double counting.
The validity of our calculations lies in the fact that these two $h$-B$_2$ sheets will be decoupled when the substrate is sufficiently thick.
The appearance of both the doubly degenerate $\sigma$-bonding bands and the nearly doubly degenerate boron phonon modes indicates that the interaction between these two $h$-B$_2$ sheets is extremely weak.
To further justify the decoupling assumption made in our current slab model, we also calculated the band structure and phonon spectrum for a thicker slab with 19 Al layers and two $h$-B$_2$ sheets.
After relaxation, the substrate is as thick as 40.65 {\AA}. We find that the $\sigma$-bonding bands of $h$-B$_2$ are entirely unchanged.
For strongly coupled phonons modes at the $\Gamma$ point, the frequency of $E_g$ modes has the largest relative deviation, whose value is only about 0.11\% in comparison to the result of the current slab model.
Hence, the coupling between these two $h$-B$_2$ sheets in current slab model can be ignored.

The EPC of $h$-B$_2$ behaves differently from that in MgB$_2$, although the metallic
$\sigma$-bonding bands of $h$-B$_2$ are preserved in $h$-B$_2$/Al(111).
At the $\Gamma$ point, detailed analysis shows that the EPC matrix elements of the bond-stretching modes ($E_u$ and $E_g$) in $h$-B$_2$ are indeed comparable to those of $E_{2g}$ modes in MgB$_2$.
However, the phonon DOS per $h$-B$_2$ near the $E_g$ modes is compressed with respect to that around the $E_{2g}$ modes in MgB$_2$ \cite{Bohnen-PRL86}, due to
well-separated out-of-plane and in-plane boron phonons [Fig.~\ref{fig:phonon}(c)].
Moreover, the frequencies of $E_u$ and $E_g$ modes at the $\Gamma$ point are higher than the $E_{2g}$ mode in MgB$_2$ (70.8 meV) \cite{Bohnen-PRL86}.
These two effects reduce the EPC in $h$-B$_2$.

Motivated by the existence of a strong coupling between boron and Al atoms, we calculated the EPC in a monolayer structured AlB$_2$, including a $h$-B$_2$ and a single Al layer.
Figure \ref{fig:a2F}(c) shows the calculated spectrum of $\alpha^2F(\omega)$.
In comparison with $h$-B$_2$/Al(111), the peak of $\alpha^2F(\omega)$ arisen from the Al phonons shifts from 35 meV to a lower frequency in this monolayer AlB$_2$.
The high-frequency peaks of $\alpha^2F(\omega)$ in monolayer AlB$_2$ are also red-shifted, because the lattice constant for monolayer AlB$_2$ is 2.5\% larger than that of $h$-B$_2$/Al(111).
As a consequence, a dramatic enhancement in $\lambda$ (1.01) and a reduction in $\omega_{\text{log}}$ (17.82 meV) are obtained.

The Coulomb pseudopotential $\mu^*$ is commonly used as a free parameter to fit experimental $T_c$.
To predict the $T_c$ for the monolayer AlB$_2$ more accurately, we deduce the value of $\mu^*$ from the bulk AlB$_2$.
The spectrum of $\alpha^2F(\omega)$ for the bulk AlB$_2$ generated by Wannier interpolation is shown in Fig.~\ref{fig:a2F}(d), consistent with the result presented in Ref.~\cite{Bohnen-PRL86}.
$\lambda$ and $\omega_{\text{log}}$ for AlB$_2$ are estimated to be 0.46 and 45.30 meV, respectively.
There would be no superconductivity in the bulk AlB$_2$ if $\mu^*$ is set to 0.23.
Assuming this is also the value of $\mu^*$, we find the $T_c$ of the monolayer AlB$_2$ will be about 6.5 K.

Compared with bulk AlB$_2$, the superconductivity in monolayer AlB$_2$ is closely related to the distance ($h_{\text{Al-B}}$) between Al and honeycomb boron sheet,
with $h_{\text{Al-B}}$=1.37 {\AA} in monolayer AlB$_2$, and 1.63 {\AA} in bulk AlB$_2$ \cite{Loa-PRB66}.
The correlation between $T_c$ and such distance also exists in graphite
intercalated compounds. For instance, $h_{\text{Ba-C}}$=2.62 {\AA}, $h_{\text{Sr-C}}$=2.47 {\AA}, and $h_{\text{Ca-C}}$=2.26 {\AA} \cite{Calandra-PRB74} in nonsuperconducting
BaC$_6$ \cite{Kim-PRL99}, SrC$_6$ ($T_c$=1.65 K) \cite{Kim-PRL99}, and CaC$_6$ ($T_c$=11.5 K) \cite{Emery-PRL95,Weller-NatPhys1}, respectively.
Now, monolayer AlB$_2$ and fcc Al are two superconductors, with $T_c$ being 6.5 K and 1.2 K, respectively. If we splice these two compounds together along the [111] direction of fcc Al, the structure of $h$-B$_2$/Al(111) will reappear exactly.
In that sense, analogous superconductivity in $h$-B$_2$/Al(111) can be anticipated in experiment, as in monolayer AlB$_2$.

\section{V.  Summary}

Based on the first-principles density functional theory and the state-of-the-art Wannier interpolation technique, we calculate the EPC for $h$-B$_2$ grown on an Al(111) substrate.
We simulate the substrate effect using a slab model.
Similar as in MgB$_2$, we find that the $\sigma$-bonding bands are also metallized
in $h$-B$_2$/Al(111).
However, the EPC constant $\lambda$ in $h$-B$_2$ is significantly smaller than
in MgB$_2$, due to the blue shift of the bond-stretching boron phonons and
the reduction of phonon DOS.
Furthermore, we find that the interlayer coupling between Al-associated phonons and electrons on the $h$-B$_2$ is quite strong.
Based on this observation, we predict that a borophene decorated by just one Al layer,
{\it i.e.} monolayer AlB$_2$, could become a superconductor below 6.5 K.
We also suggest that similar superconducting $T_c$ can be detected in $h$-B$_2$/Al(111).

\section{acknowledgments}
This work is supported by the National Key R \& D Program of China
(Grant No. 2017YFA0302900),
National Natural Science Foundation of China (Grant Nos. 11774422, and 11674185),
and Zhejiang Provincial
Natural Science Foundation of China under Grant No. LY17A040005.
M.G. is also sponsored by K.C.Wong Magna Fund in Ningbo University.

\appendix

\section{Appendix A: borophene on semi-infinite Al substrate}

With the reasonable EPC constant of $h$-B$_2$ in hand, we further employed another geometry for the slab model with borophene layer on the semi-infinite Al substrate,
to see how much it deviates from the obtained results.
For clarity, we label this slab as $h$-B$_2$/Al(111)$_\text{SI}$ to distinguish it from the $h$-B$_2$/Al(111) in the main text.
If there exists rich dangling bonds on the fracture surface (e.g. silicon surface), the dangling bonds should be passivated with hydrogen.
However, the passivation strategy is not suitable for Al surface, especially when studying the EPC. Firstly, Al is a good metal, there is not
covalent-type dangling bonds on its surface. Secondly, passivation with hydrogen will introduce extra high-frequency phonon modes.
In our calculations, the semi-infinite substrate was represented by four layers of Al, in which the bottom two layers were fixed to simulate
the bulk properties. The top two layers and the borophene sheet were allowed to relax freely. The $c$-axis thickness of the slab was set to 20 {\AA}.

\begin{figure}[tbh]
\begin{center}
\includegraphics[width=8.6cm]{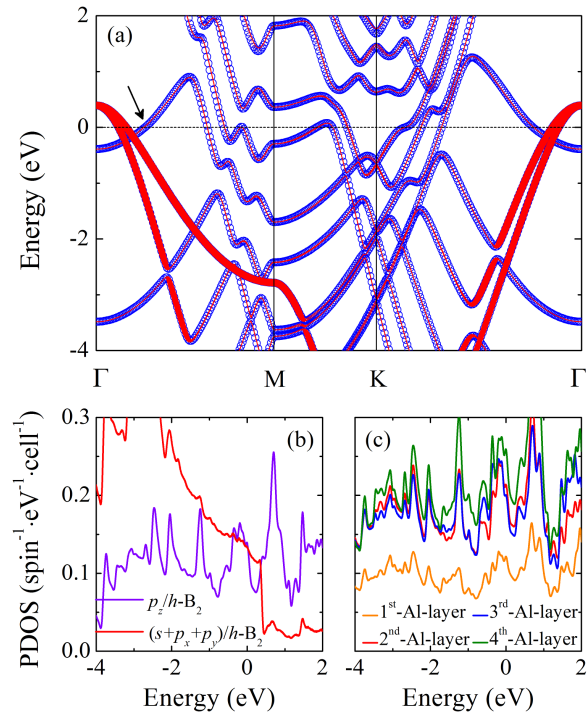}
\caption{(a) Band structures, (b) PDOS of boron orbitals, and (c) PDOS of Al layers for $h$-B$_2$/Al(111)$_\text{SI}$. The meanings of symbols are the same as in Fig.~\ref{fig:Band}.
Here, the 4$^{th}$-Al-layer corresponds to the fracture surface of semi-infinite Al substrate.}
\label{fig:Band-SI}
\end{center}
\end{figure}

\begin{figure}[tbh]
\begin{center}
\includegraphics[width=8.6cm]{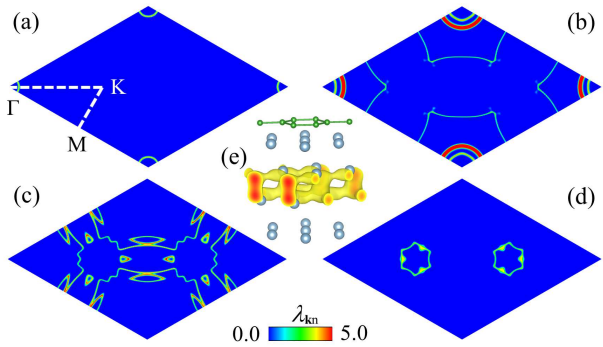}
\caption{(a)-(d) Fermi surfaces weighted by $\lambda_{\bf{k}\nu}$ in $h$-B$_2$/Al(111)$_\text{SI}$. (e) Charge density distribution of strongly coupled electronic state pointed by black arrow in Fig.~\ref{fig:Band-SI}(a).
The isovalue for charge density is set to 5.0$\times$10$^{-3}$ $e$/Bohr$^3$.}
\label{fig:FS-SI}
\end{center}
\end{figure}

After optimization, the buckling height of $h$-B$_2$ is found to be 0.031 {\AA} in $h$-B$_2$/Al(111)$_\text{SI}$, close to that in $h$-B$_2$/Al(111).
The band structures and PDOS are shown in Fig.~\ref{fig:Band-SI}. Although the band structures differ from $h$-B$_2$/Al(111), the $\sigma$ bands are
almost unaffected [Fig.~\ref{fig:Band-SI}(a) and Fig.~\ref{fig:Band-SI}(b)]. For example, the $\sigma$-band maxima at the $\Gamma$ point are equal to 0.39 eV and 0.40 eV for $h$-B$_2$/Al(111)$_\text{SI}$ and $h$-B$_2$/Al(111), respectively.
Significant charge transfer from the outmost Al layer to borophene can be inferred from Fig.~\ref{fig:Band-SI}(c).
The DOS for the 4$^{th}$ Al layer is slightly larger than other Al layers, suggesting the existence of surface states.
There are four bands crossing the Fermi level, giving rise to four Fermi surfaces [Fig.~\ref{fig:FS-SI}].
The small hole pockets surrounding the $\Gamma$ point in Fig.~\ref{fig:FS-SI}(a) and Fig.~\ref{fig:FS-SI}(b) are resulted from the $\sigma$ bands.
The electronic states on the red circle near the $\Gamma$ point dominate the contribution to EPC of $h$-B$_2$/Al(111)$_\text{SI}$.
However, charge density analysis indicates that the strongly coupled electronic states are from Al substrate, not borophene [Fig.~\ref{fig:FS-SI}(e)].

\begin{figure}[tbh]
\begin{center}
\includegraphics[width=8.6cm]{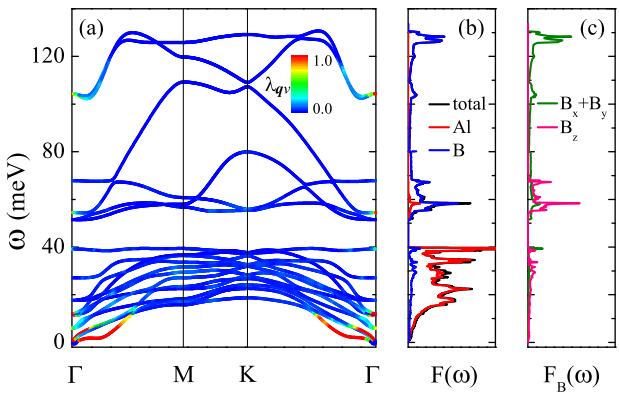}
\caption{(a) $\lambda_{\bf{q}\nu}$-weighted phonon spectrum, (b) projected phonon density of states, and (c) decomposed boron-related phonon density of states along three directions for $h$-B$_2$/Al(111)$_\text{SI}$.}
\label{fig:phonon-SI}
\end{center}
\end{figure}

\begin{figure}[tbh]
\begin{center}
\includegraphics[width=8.6cm]{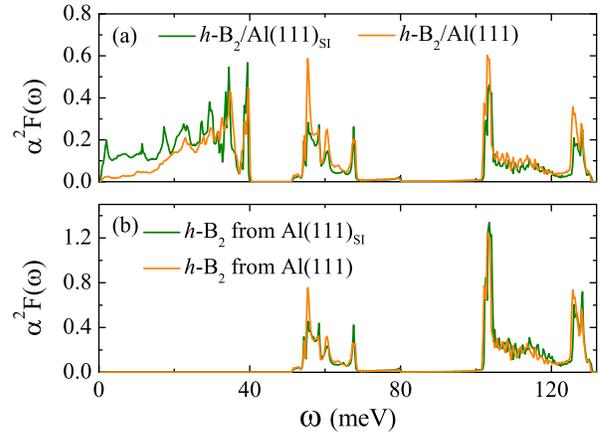}
\caption{Eliashberg spectral function $\alpha^2F(\omega)$ for (a) $h$-B$_2$/Al(111)$_\text{SI}$, (b) $h$-B$_2$ extracted from $h$-B$_2$/Al(111)$_\text{SI}$.
These two quantities for $h$-B$_2$/Al(111) are also given to make a comparison.}
\label{fig:a2F-SI}
\end{center}
\end{figure}

The $\lambda_{\bf{q}\nu}$-weighted phonon spectrum and phonon DOS in $h$-B$_2$/Al(111)$_\text{SI}$ are shown in Fig.~\ref{fig:phonon-SI}.
The phonon modes associated with boron are almost the same as in Fig.~\ref{fig:phonon}(a). Boron and Al phonons are still well separated [Fig.~\ref{fig:phonon-SI}(b)]. The lowest acoustic branch exhibits obvious softening. This results in large EPC constant $\lambda_{\bf{q}\nu}$ [Fig.~\ref{fig:phonon-SI}(a)].
Compared with $h$-B$_2$/Al(111), the low-frequency part of $\alpha^2F(\omega)$ for $h$-B$_2$/Al(111)$_\text{SI}$ is greatly enhanced [Fig.~\ref{fig:a2F-SI}(a)], self-consistent with the softened acoustic phonons. Thus strong EPC and small $\omega_{\text{log}}$ will be obtained.
As expected, we find that the EPC constant and $\omega_{\text{log}}$ are 1.26 and 8.27 meV, respectively. The
$\alpha^2F(\omega)$ of $h$-B$_2$ can also be extracted from $h$-B$_2$/Al(111)$_\text{SI}$ following the procedure in the main text.
Similarly, we find that the $\sigma$-bonding electrons of $h$-B$_2$ contribute 38.1\% (68.5\%) to the EPC caused by the out-of-plane (in-plane) boron phonons.
As a result, the $\alpha^2F(\omega)$ in the intervals [50, 100] meV and [100, 140] meV should be reduced by 61.9\% and 31.5\%, respectively.
The $N(0)$ and $N(0)$ contributed by the $h$-B$_2$ electrons are 1.15 and 0.27 states/spin/eV/cell. The refined $\alpha^2F(\omega)$ of $h$-B$_2$ is shown in Fig.~\ref{fig:a2F-SI}(b).
After integrating $\alpha^2F(\omega)$, we find $\lambda$ of $h$-B$_2$ is 0.23, very close to the value determined in the main text.
As far as the intrinsic EPC of $h$-B$_2$ is concerned, the fracture surface of semi-infinite Al substrate has negligible effect.
Thus determining the EPC of $h$-B$_2$ with semi-infinite Al substrate is also a feasible scheme with reduced computational cost.


\begin{references}
\bibitem{Yang-PRB77}X. Yang, Y. Ding, and J. Ni,
\href{https://dx.doi.org/10.1103/PhysRevB.77.041402}{Phys. Rev. B {\bf 77}, 041402(R) (2008)}.

\bibitem{Penev-NanoLett12}E. S. Penev, S. Bhowmick, A. Sadrzadeh, and B. I. Yakobson,
\href{https://dx.doi.org/10.1021/nl3004754}{Nano Lett. {\bf 12}, 2441 (2012)}.

\bibitem{Yu-JPCC116}X. Yu, L. Li, X.-W. Xu, and C.-C. Tang,
\href{http://dx.doi.org/10.1021/jp305545z}{J. Phys. Chem. C {\bf 116}, 20075 (2012)}.

\bibitem{Tang-PRB82}H. Tang and S. Ismail-Beigi,
\href{https://dx.doi.org/10.1103/PhysRevB.82.115412}{Phys. Rev. B {\bf 82}, 115412 (2010)}.

\bibitem{Wu-ACSNano6}X. Wu, J. Dai, Y. Zhao, Z. Zhuo, J. Yang, and X. C. Zeng,
\href{https://dx.doi.org/10.1021/nn302696v}{ACS Nano {\bf 6}, 7443 (2012)}.

\bibitem{Mannix-Science}A. J. Mannix, X.-F. Zhou, B. Kiraly, J. D. Wood, D. Alducin, B. D. Myers, X. Liu, B. L. Fisher, U. Santiago, J. R. Guest, M. J. Yacaman, A. Ponce, A. R. Oganov, M. C. Hersam, and N. P. Guisinger,
\href{https://dx.doi.org/10.1126/science.aad1080}{Science {\bf 350}, 1513 (2015)}.

\bibitem{Feng-arXiv1}B. Feng, J. Zhang, Q. Zhong, W. Li, S. Li, H. Li, P. Cheng, S. Meng, L. Chen, and K. Wu,
\href{https://dx.doi.org/10.1038/nchem.2491}{Nat. Chem. {\bf 8}, 563 (2016)}.

\bibitem{Franceschi-Nature5}S. De Franceschi, L. Kouwenhoven, Ch. Sch\"{o}nenberger, and W. Wernsdorfer,
\href{https://dx.doi.org/10.1038/nnano.2010.173}{Nat. Nanotechnol. {\bf 5}, 703 (2010)}.

\bibitem{Huefner-PRB79}M. Huefner, C. May, S. Ki\v{c}in, K. Ensslin, T. Ihn, M. Hilke, K. Suter, N. F. de Rooij, and U. Staufer,
\href{https://dx.doi.org/10.1103/PhysRevB.79.134530}{Phys. Rev. B {\bf 79}, 134530 (2009)}.

\bibitem{Liu-Nano8}H. Liu, A.T. Neal, Z. Zhu, Z. Luo, X. Xu, D. Tom\'{a}nek, and P. D. Ye,
\href{https://dx.doi.org/10.1021/nn501226z}{ACS Nano {\bf 8}, 4033 (2014)}.

\bibitem{Li-Nature9}L. Li, Y. Yu, G. J. Ye, Q. Ge, Xe. Ou, H. Wu, D. Feng, X. H. Chen, and Y. Zhang,
\href{https://dx.doi.org/10.1038/nnano.2014.35}{Nat. Nanotechnol. {\bf 9}, 372 (2014)}.

\bibitem{Lalmi-APL97}B. Lalmi, H. Oughaddou, H. Enriquez, A. Kara, S. Vizzini, B. Ealet, and B. Aufray,
\href{https://dx.doi.org/10.1063/1.3524215}{Appl. Phys. Lett. {\bf 97}, 223109 (2010)}.

\bibitem{Chen-PRL109}L. Chen, C.-C. Liu, B. Feng, X. He, P. Cheng, Z. Ding, S. Meng, Y. Yao, and K. Wu,
\href{https://doi.org/10.1103/PhysRevLett.109.056804}{Phys. Rev. Lett. {\bf 109}, 056804 (2012)}.

\bibitem{Davila-NJP16}M. E. D\'{a}vila, L. Xian, S. Cahangirov, A. Rubio, and G. Le Lay,
\href{https://doi.org/10.1088/1367-2630/16/9/095002}{New J. Phys. {\bf 16}, 095002 (2014)}.

\bibitem{Derivaz-NanoLett15}M. Derivaz, D. Dentel, R. Stephan, M.-C. Hanf, A. Mehdaoui, P. Sonnet, and C. Pirri,
\href{http://dx.doi.org/10.1021/acs.nanolett.5b00085}{Nano Lett. {\bf 15}, 2510 (2015)}.

\bibitem{Zhu-NatMat14}F.-F. Zhu, W.-J. Chen, Y. Xu, C.-L. Gao, D.-D. Guan, C.-H. Liu, D. Qian, S.-C. Zhang, and J.-F. Jia,
\href{https://doi.org/10.1038/nmat4384}{Nat. Mater. {\bf 14}, 1020 (2015)}.

\bibitem{Zang-arXiv}Y. Zang, T. Jiang, Y. Gong, Z. Guan, C. Liu, M. Liao, K. Zhu, Z. Li, L. Wang, W. Li, C. Song, D. Zhang, Y. Xu, K. He, X. Ma, S.-C. Zhang, and Q.-K. Xue,
\href{https://doi.org/10.1002/adfm.201802723}{Adv. Funct. Mater. {\bf 28}, 1802723 (2018)}.

\bibitem{Qiao-NatCommun5}J. Qiao, X. Kong, Z.-X. Hu, F. Yang, and W. Ji,
\href{https://doi.org/10.1038/ncomms5475}{Nat. Commun. {\bf 5}, 4475 (2014)}.

\bibitem{Cahangirov-PRL102}S. Cahangirov, M. Topsakal, E. Akt\"{u}rk, H. \c{S}ahin, and S. Ciraci,
\href{https://doi.org/10.1103/PhysRevLett.102.236804}{Phys. Rev. Lett. {\bf 102}, 236804 (2009)}.

\bibitem{Xu-PRL111}Y. Xu, B. Yan, H.-J. Zhang, J. Wang, G. Xu, P. Tang, W. Duan, and S.-C. Zhang,
\href{https://doi.org/10.1103/PhysRevLett.111.136804}{Phys. Rev. Lett. {\bf 111}, 136804 (2013)}.

\bibitem{Profeta-Nature8}G. Profeta, M. Calandra, and F. Mauri,
\href{https://dx.doi.org/10.1038/nphys2181}{Nat. Phys. {\bf 8}, 131 (2012)}.

\bibitem{Wan-EPL104}W. Wan, Y. Ge, F. Yang, and Y. Yao,
\href{https://dx.doi.org/10.1209/0295-5075/104/36001}{Europhys. Lett. {\bf 104}, 36001 (2013)}.

\bibitem{Shao-EPL108}D. F. Shao, W. J. Lu, H. Y. Lv, and Y. P. Sun,
\href{https://dx.doi.org/10.1209/0295-5075/108/67004}{Europhys. Lett. {\bf 108}, 67004 (2014)}.

\bibitem{Ge-NJP17}Y. Ge, W. Wan, F. Yang, and Y. Yao,
\href{https://dx.doi.org/10.1088/1367-2630/17/3/035008}{New J. Phys. {\bf 17}, 035008 (2015)}.

\bibitem{Si-PRL111}C. Si, Z. Liu, W. Duan, and F. Liu,
\href{https://doi.org/10.1103/PhysRevLett.111.196802}{Phys. Rev. Lett. {\bf 111}, 196802 (2013)}.

\bibitem{Kong-CPB27}X. Kong, M. Gao, X.-W. Yan, Z.-Y. Lu, and T. Xiang,
\href{https://doi.org/10.1088/1674-1056/27/4/046301}{Chin. Phys. B {\bf 27}, 046301 (2018)}.

\bibitem{Tiwari-arXiv}A. P. Tiwari, S. Shin, E. Hwang, S.-G. Jung, T. Park, and H. Lee,
\href{http://arxiv.org/abs/1508.06360}{arXiv:1508.06360}.

\bibitem{Cao-Nature556}Y. Cao, V. Fatemi, S. Fang, K. Watanabe, T. Taniguchi, E. Kaxiras, and P. Jarillo-Herrero,
\href{https://doi.org/10.1038/nature26160}{Nature (London) {\bf 556}, 43 (2018)}.

\bibitem{Liao-arXiv}M. Liao, Y. Zang, Z. Guan, H. Li, Y. Gong, K. Zhu, X.-P. Hu, D. Zhang, Y. Xu, Y.-Y. Wang, K. He, X.-C. Ma, S.-C. Zhang, and Q.-K. Xue,
\href{https://doi.org/10.1038/s41567-017-0031-6}{Nat. Phys. {\bf 14}, 344 (2018)}.

\bibitem{Penev-Nano16}E. S. Penev, A. Kutana, and B. I. Yakobson,
\href{https://doi.org/10.1021/acs.nanolett.6b00070}{Nano Lett. {\bf 16}, 2522 (2016)}.

\bibitem{Zhao-APL108}Y. Zhao, S. Zeng, and Jun Ni,
\href{https://doi.org/10.1063/1.4953775}{Appl. Phys. Lett. {\bf 108}, 242601 (2016)}.

\bibitem{Gao-PRB95}M. Gao, Q.-Z. Li, X.-W. Yan, and J. Wang,
\href{https://doi.org/10.1103/PhysRevB.95.024505}{Phys. Rev. B {\bf 95}, 024505 (2017)}.

\bibitem{Xiao-APL109}R. C. Xiao, D. F. Shao, W. J. Lu, H. Y. Lv, J. Y. Li, and Y. P. Sun,
\href{https://doi.org/10.1063/1.4963179}{Appl. Phys. Lett. {\bf 109}, 122604 (2016)}.

\bibitem{Cheng-2D4}C. Cheng, J.-T. Sun, H. Liu, H.-X. Fu, J. Zhang, X.-R. Chen, and S. Meng,
\href{https://doi.org/10.1088/2053-1583/aa5e1b}{2D Mater. {\bf 4}, 025032 (2017)}.

\bibitem{Somoano-PRL27}R. B. Somoano and A. Rembaum,
\href{https://doi.org/10.1103/PhysRevLett.27.402}{Phys. Rev. Lett. {\bf 27}, 402 (1971)}.

\bibitem{Morosan-NP2}E. Morosan, H. W. Zandbergen, B. S. Dennis, J. W. G. Bos,
Y. Onose, T. Klimczuk, A. P. Ramirez, N. P. Ong, and R. J. Cava,
\href{https://doi.org/10.1038/nphys360}{Nat. Phys. {\bf 2}, 544 (2006)}.

\bibitem{Li-Nature529} L. J. Li, E. C. T. O'Farrell, K. P. Loh, G. Eda, B. \"{O}zyilmaz, and A. H. Castro,
\href{https://doi.org/10.1038/nature16175}{Nature (London) {\bf 529}, 185 (2016)}.

\bibitem{Yu-NN10} Y. J. Yu, F. Yang, X. F. Lu, Y. J. Yan, Y.-H. Cho, L. Ma, X. Niu, S. Kim, Y.-W. Son, D. Feng, S. Li, S.-W. Cheong, X. H. Chen, and Y. Zhang,
\href{https://doi.org/10.1038/nnano.2014.323}{Nat. Nanotechnol. {\bf 10}, 270 (2015)}.

\bibitem{Ye-Science338}J. T. Ye, Y. J. Zhang, R. Akashi, M. S. Bahramy, R. Arita, and Y. Iwasa,
\href{https://doi.org/10.1126/science.1228006}{Science {\bf 338}, 1193 (2012)}.

\bibitem{Costanzo-NN11}D. Costanzo, S. Jo, H. Berger, and A. F. Morpurgo,
\href{https://doi.org/10.1038/nnano.2015.314}{Nat. Nanotechnol. {\bf 11}, 339 (2016)}.

\bibitem{Shi-SR5}W. Shi, J. Ye, Y. Zhang, R. Suzuki, M. Yoshida, J. Miyazaki, N. Inoue, Y. Saito, and Y. Iwasa,
\href{https://doi.org/10.1038/srep12534}{Sci. Rep. {\bf 5}, 12534 (2015)}.

\bibitem{Chi-PRL120}Z. Chi, X. Chen, F. Yen, F. Peng, Y. Zhou, J. Zhu, Y. Zhang, X. Liu, C. Lin, S. Chu, Y. Li, J.Zhao, T. Kagayama, Y. Ma, and Z. Yang,
\href{https://doi.org/10.1103/PhysRevLett.120.037002}{Phys. Rev. Lett. {\bf 120}, 037002 (2018)}.

\bibitem{Revolinsky-JPCS26}E. Revolinsky, G. A. Spiering, and D. J. Beerntsen,
\href{https://doi.org/10.1016/0022-3697(65)90190-3}{J. Phys. Chem. Solids {\bf 26}, 1029 (1965)}.

\bibitem{Valla-PRL85}T. Valla, A. V. Fedorov, P. D. Johnson, J. Xue, K. E. Smith, and F. J. DiSalvo,
\href{https://doi.org/10.1103/PhysRevLett.85.4759}{Phys. Rev. Lett. {\bf 85}, 4759 (2000)}.

\bibitem{Freitas-PRB93}D. C. Freitas, P. Rodi\`{e}re, M. R. Osorio, E. Navarro-Moratalla, N. M. Nemes, V. G. Tissen, L. Cario, E. Coronado, M. Garc\'{\i}a-Hern\'{a}ndez, S. Vieira, M. N\'{u}\~{n}ez-Regueiro, and H. Suderow,
\href{https://doi.org/10.1103/PhysRevB.93.184512}{Phys. Rev. B {\bf 93}, 184512 (2016)}.

\bibitem{Nagata-JPCS53}S. Nagata, T. Aochi, T. Abe, S. Ebisu, T. Hagino, Y. Seki, K. Tsutsumi,
\href{https://doi.org/10.1016/0022-3697(92)90242-6}{J. Phys. Chem. Solids {\bf 53}, 1259 (1992)}.

\bibitem{Castro-PRL86}A. H. Castro Neto,
\href{https://doi.org/10.1103/PhysRevLett.86.4382}{Phys. Rev. Lett. {\bf 86}, 4382 (2001)}.

\bibitem{Guillamon-NJP13}I. Guillam\'{o}n, H. Suderow, J. G. Rodrigo, S. Vieira, P. Rodi\`{e}re,
L. Cario, E. Navarro-Moratalla, C. Mart\'{\i}-Gastaldo, and E. Coronado,
\href{https://doi.org/10.1088/1367-2630/13/10/103020}{New J. Phys. {\bf 13}, 103020 (2011)}.

\bibitem{Navarro-NC7}Efr\'{e}n Navarro-Moratalla, J. O. Island, S. Ma\~{n}as-Valero, E. Pinilla-Cienfuegos, A. Castellanos-Gomez, J. Quereda, G. Rubio-Bollinger, L. Chirolli, J. A. Silva-Guill\'{e}n,
N. Agra\"{\i}t, G. A. Steele, F. Guinea, H. S. J. van der Zant, and E. Coronado,
\href{https://doi.org/10.1038/ncomms11043}{Nat. Commun. {\bf 7}, 11043 (2016)}.

\bibitem{Yang-PRB98}Y. Yang, S. Fang, V. Fatemi, J. Ruhman, E. Navarro-Moratalla, K. Watanabe, T. Taniguchi, E. Kaxiras, and P. Jarillo-Herrero,
\href{https://doi.org/10.1103/PhysRevB.98.035203}{Phys. Rev. B {\bf 98}, 035203 (2018)}.

\bibitem{Xi-NP12}X. Xi, Z. Wang, W. Zhao, J.-H. Park, K. T. Law, H. Berger, L. Forr\'{o}, J. Shan, and K. F. Mak,
\href{https://doi.org/10.1038/nphys3538}{Nat. Phys. {\bf 12}, 139 (2016)}.

\bibitem{Xi-NN10}X. Xi, L. Zhao, Z. Wang, H. Berger, L. Forr\'{o}, J. Shan, and K. F. Mak,
\href{https://doi.org/10.1038/nnano.2015.143}{Nat. Nanotechnol. {\bf 10}, 765 (2015)}.

\bibitem{Li-arXiv1712}W. Li, L. Kong, C. Chen, J. Gou, S. Sheng, W. Zhang, H. Li, L. Chen, P. Cheng, and K. Wu,
\href{https://doi.org/10.1016/j.scib.2018.02.006}{Sci. Bull. {\bf 63}, 282 (2018)}.

\bibitem{MgB2} J. Nagamatsu, N. Nakagawa, T. Muranaka, Y. Zenitani, and J. Akimitsu,
\href{https://doi.org/10.1038/35065039}{Nature (London) {\bf 410}, 63 (2001)}.

\bibitem{Margine-PRB87}E. R. Margine and F. Giustino,
\href{https://doi.org/10.1103/PhysRevB.87.024505}{Phys. Rev. B {\bf 87}, 024505 (2013)}.

\bibitem{pwscf}P. Giannozzi {\it et al.},
\href{https://doi.org/10.1088/0953-8984/21/39/395502}{J. Phys.: Condens. Matter {\bf 21}, 395502 (2009)}.

\bibitem{Giustino-PRB76}F. Giustino, M. L. Cohen, and S. G. Louie,
\href{https://doi.org/10.1103/PhysRevB.76.165108}{Phys. Rev. B {\bf 76}, 165108 (2007)}.

\bibitem{Methfessel-PRB40}M. Methfessel and A. T. Paxton,
\href{https://doi.org/10.1103/PhysRevB.40.3616}{Phys. Rev. B {\bf 40}, 3616 (1989)}.

\bibitem{Baroni-RMP73_515}S. Baroni, S. de Gironcoli, A. Dal Corso, and P. Giannozzi,
\href{https://doi.org/10.1103/RevModPhys.73.515}{Rev. Mod. Phys. {\bf 73}, 515 (2001)}.

\bibitem{Wyckoff}R.W.G. Wyckoff, Crystal Structure {\bf 1}, 7-83, Second edition. Interscience Publishers, New York (1963).

\bibitem{Marzari-PRB56}N. Marzari and D. Vanderbilt,
\href{https://doi.org/10.1103/PhysRevB.56.12847}{Phys. Rev. B {\bf 56}, 12847 (1997)}.

\bibitem{Souza-PRB65}I. Souza, N. Marzari, and D. Vanderbilt,
\href{https://doi.org/10.1103/PhysRevB.65.035109}{Phys. Rev. B {\bf 65}, 035109 (2001)}.

\bibitem{Mostofi-CPC178}A. A. Mostofi, J. R. Yates, Y.-S. Lee, I. Souza, D. Vanderbilt, and N. Marzari,
\href{https://doi.org/10.1016/j.cpc.2007.11.016}{Comput. Phys. Commun. {\bf 178}, 685 (2008)}.

\bibitem{Noffsinger-CPC181}J. Noffsinger, F. Giustino, B. D. Malone, C.-H. Park, S. G. Louie, and M. L. Cohen,
\href{https://doi.org/10.1016/j.cpc.2010.08.027}{Comput. Phys. Commun. {\bf 181}, 2140 (2010)}.

\bibitem{Allen-PRB6_2577}P. B. Allen,
\href{https://doi.org/10.1103/PhysRevB.6.2577}{Phys. Rev. B {\bf 6}, 2577 (1972)}.

\bibitem{Allen-RPB12_905}P. B. Allen and R. C. Dynes,
\href{https://doi.org/10.1103/PhysRevB.12.905}{Phys. Rev. B {\bf 12}, 905 (1975)}.

\bibitem{Margine-PRB90}E. R. Margine and F. Giustino,
\href{10.1103/PhysRevB.90.014518}{Phys. Rev. B {\bf 90}, 014518 (2014)}.

\bibitem{Medvedeva-PRB64}N. I. Medvedeva, A. L. Ivanovskii, J. E. Medvedeva and A. J. Freeman,
\href{https://doi.org/10.1103/PhysRevB.64.020502}{Phys. Rev. B {\bf 64}, 020502(R) (2001)}.

\bibitem{Henkelman-CMS36} G. Henkelman, A. Arnaldsson, and H. J\'{o}nsson,
\href{https://doi.org/10.1016/j.commatsci.2005.04.010}{Comput. Mater. Sci. {\bf 36}, 354 (2006)}.

\bibitem{Tang-JPCM21}W. Tang, E. Sanville, and G. Henkelman,
\href{https://doi.org/10.1088/0953-8984/21/8/084204}{J. Phys.: Condens. Matter {\bf 21}, 084204 (2009)}.

\bibitem{Savrasov-PRB54}S. Y. Savrasov and D. Y. Savrasov,
\href{https://doi.org/10.1103/PhysRevB.54.16487}{Phys. Rev. B {\bf 54}, 16487 (1996)}.

\bibitem{Bohnen-PRL86}K.-P. Bohnen, R. Heid, and B. Renker,
\href{https://doi.org/10.1103/PhysRevLett.86.5771}{Phys. Rev. Lett. {\bf 86}, 5771 (2001)}.

\bibitem{Loa-PRB66}I. Loa, K. Kunc, K. Syassen, and P. Bouvier,
\href{https://doi.org/10.1103/PhysRevB.66.134101}{Phys. Rev. B {\bf 66}, 134101 (2002)}.

\bibitem{Calandra-PRB74}M. Calandra and F. Mauri,
\href{https://doi.org/10.1103/PhysRevB.74.094507}{Phys. Rev. B {\bf 74}, 094507 (2006)}.

\bibitem{Kim-PRL99}J. S. Kim, L. Boeri, J. R. O'Brien, F. S. Razavi, and R. K. Kremer,
\href{https://doi.org/10.1103/PhysRevLett.99.027001}{Phys. Rev. Lett. {\bf 99}, 027001 (2007)}.

\bibitem{Emery-PRL95}N. Emery, C. H\'{e}rold, M. d'Astuto, V. Garcia, Ch. Bellin, J. F. Mar\^{e}ch\'{e}, P. Lagrange, and G. Loupias,
\href{https://doi.org/10.1103/PhysRevLett.95.087003}{Phys. Rev. Lett. {\bf 95}, 087003 (2005)}.

\bibitem{Weller-NatPhys1}T. E. Weller, M. Ellerby, S. S. Saxena, R. P. Smith, and N. T. Skipper,
\href{https://doi.org/10.1038/nphys0010}{Nat. Phys. {\bf 1}, 39 (2005)}.

\end{references}
\end{document}